\documentclass{emulateapj}
\makeatletter

\newcommand{\Rmnum}[1]{\expandafter\@slowromancap\romannumeral #1@}
\makeatother

\slugcomment{Astronomical Journal, manuscript.}

\shorttitle{The scattering outcomes of \emph{Kepler} circumbinary planets}
\shortauthors{Gong \& Ji}

\begin{document}

\title{The scattering outcomes of \emph{Kepler} circumbinary planets: planet mass ratio}

\author{Yan-Xiang Gong\altaffilmark{1,2}, Jianghui Ji\altaffilmark{1}}

\affil{$^{1}$ CAS Key Laboratory of Planetary Sciences, Purple Mountain Observatory, Chinese Academy of Sciences, Nanjing 210008, China}

\affil{$^{2}$ College of Physics and Electronic Engineering, Taishan
University, Taian 271000, China}

\email{yxgong@pmo.ac.cn, jijh@pmo.ac.cn}

\begin{abstract}
Recent studies reveal that the free eccentricities of Kepler-34b and Kepler-413b are much larger than their forced eccentricities, implying that the scattering events may take place in their formation. The observed orbital configuration of Kepler-34b cannot be well reproduced in disk-driven migration models, whereas a two-planet scattering scenario can play a significant role of shaping the planetary configuration. These studies indicate that circumbinary planets discovered by \emph{Kepler} may have experienced scattering process. In this work, we extensively investigate the scattering outcomes of circumbinary planets focusing on the effects of \emph{planet mass ratio}. We find that the planetary mass ratio and the the initial relative locations of planets act as two important parameters which affect the eccentricity distribution of the surviving planets. As an application of our model, we discuss the observed orbital configurations of Kepler-34b and Kepler-413b. We first adopt the results from the disk-driven models as the initial conditions, then simulate the scattering process occurred in the late evolution stage of circumbinary planets. We show that the present orbital configurations of Kepler-34b and Kepler-413b can be  well reproduced when considering two unequal-mass planet \emph{ejection} model. Our work further suggests that some of the currently discovered circumbinary single-planet systems may be the survivals of original multiple-planet systems. The disk-driven migration and the scattering events occurring in the late stage both play an irreplaceable role in sculpting the final systems.
\end{abstract}

\keywords{Methods: numerical --- planets and satellites: dynamical evolution and stability --- planets and satellites: individual (Kepler-34b, Kepler-413b) --- binaries: close}

\section{Introduction}

So far over 3400 planets have been discovered by the \emph{Kepler} space telescope \citep{Bor2010,Lis2011,Bata13, Mazeh13, Fab14, Wang2015}. These planets show a variety of orbital configurations, which has been greatly improved our understanding of planetary formation \citep{lee13, Dong13, Jin2014, raymond2014, wang14, Batygin2016, Dong17}. For example, multiple small planets with orbital periods less than $\sim 50$ days are located in many systems known as Tightly spaced Inner Planets \citep{macdonald16}, while plenty of planets are found in or near mean motion resonances in these systems \citep{wang12, wang14, wang17, lee13, Zhang2014, marti16, mills16, sun17}.

One of the exciting findings of \emph{Kepler} is the discovery of several circumbinary planets around main-sequence stars. Due to the perturbation of the inner binary, their formation, orbital characteristics and the habitability of these special bodies bring new challenges to planetary science. At present, 11 circumbinary planets have been discovered by \emph{Kepler} belonging to 9 planetary systems ({\it http://exoplanet.eu/}). Among them, Kepler-47 is a multiple-planet system \citep{orosz12, welsh15, hinse15}.

The masses and orbital configurations of these planets are listed in Table 1. The $a_{c}$ in Table 1 is the critical Semimajor Axis (SMA) of a planet (relative to the barycenter of the binary) beyond which planetary orbits can maintain long-term stability except for the unstable islands associated with N:1 resonance with the binary \citep{holman99}.
\begin{equation}
\begin{array}{*{20}{l}}
{{{{a_c}} \mathord{\left/
 {\vphantom {{{a_c}} {{a_B}}}} \right.
 \kern-\nulldelimiterspace} {{a_B}}} = 1.60 + 5.1{e_B} - 2.22e_B^2 + 4.12\mu}\\
{\;\;\;\;\;\;\;\;\;\;\;\;\;\; - 4.27{e_B}\mu  - 5.09{\mu ^2} + 4.61e_B^2{\mu ^2},}
\end{array}
\end{equation}
\label{ac}
where $\mu  = {{{m_B}} \mathord{\left/ {\vphantom {{{m_B}}
{\left( {{m_A} + {m_B}} \right)}}}
\right.\kern-\nulldelimiterspace} {\left( {{m_A} + {m_B}}
\right)}}$ is the mass ratio of the binary, $e_B$ and $a_{B}$ are its eccentricity and SMA, respectively.
Several characteristics of these planets are noteworthy: (1) The binary and planetary orbits are aligned within a few degrees. The highest relative inclination is 2.5 degree \citep{winn15}. (2) Except Kepler-1647b, most of them cluster just outside of the zone of instability. (3) The majority of them have a nearly circular orbit. Despite some of the above trends lurks the specter of selection effects, these characteristics are consistent with the results predicted by disk-driven migration models \citep{pierens07, pierens13, kley14, kley15}.

Did these circumbinary planets undergo planet-planet scattering (PPS) processes? Recent studies provide some clues to this issue. \citet{bromley15} examined the orbital characteristics of circumbinary in the context of current planet formation scenarios. They found that the forced eccentricity at Kepler-34b's location is low (about 0.002). But the eccentricity of the observed planet is much larger ($\sim 0.18$). Such high free eccentricity and the high mass of the planet favor a scattering process. As a comparison, Kepler-413b has a significant free eccentricity about 0.12. However, its forced eccentricity is only 0.003. Large free eccentricity of Kepler-413b tends to preclude the migrate-in-gas mode \citep{bromley15}. Its orbital configuration is consistent with scattering events.

\citet{kley15} considered the disk-driven migration of Kepler-34b using two-dimensional hydrodynamical simulations. They found that the planet's final equilibrium position lies beyond the observed location of Kepler-34b. To account for the closer orbit of Kepler-34b, they proposed a scenario in which there are two planets in the system. The convergent migration of the two planets often leads to capture into mean-motion resonances (MMR).  A weak planet-planet scattering process ensues when the inner planet orbits inside the gap of the disk. The above-mentioned scenario can reproduce the observed orbit of Kepler-34b. In this model, another planet would still reside in the system on long-period orbit ($\sim 1.5$ au) which may fail to transit the central binary during the operation of \emph{Kepler} as they suggested.

In this work, we extensively explored the scattering outcomes of circumbinary planets focusing on the effects of \emph{planet mass ratio}. In a single star system, the planet mass ratio is a key parameter in dynamical scattering process \citep{ford08}. Two equal-mass planets scattering model gives a narrow distribution of final eccentricities,  which cannot reproduce the eccentricity distribution of the observed giant planets, whereas two unequal-mass planet scattering model predicts a broader range of final eccentricities. With a reasonable distribution of planet mass ratios, the observed eccentricities can be reproduced \citep{ford08, chatterjee08}. However, how does planet mass ratio affect the scattering outcomes of circumbinary planets?  In \citet{gong17}, we investigated the scattering process of two \emph{equal-mass} planets by considering the role of binary configurations (i.e. $\mu$ and $e_B$). We explored all kinds of close binary configurations and showed some new features of scattering events, which differ from the scattering results revealed in single star systems. Herein, based on two unequal-mass planet scattering model, we make an extensive study of scattering outcomes of circumbinary planets through numerical simulations, then explore the final statistical configurations of the surviving planets \emph{after} PPS, in an aim to understand the formation of currently observed \emph{Kepler} circumbinary companions.

Scattering events have been found in the hydrodynamic simulations of multiple planets in a circumbinary disc. For example, disk-driven migration of multiple low mass circumbinary planets (5-20 Earth mass) in an artificial binary system were discussed in \citet{pierens08}. They showed that two planets usually undergo dynamical scattering for mass ratio $q = m_{inner}/m_{outer} < 1$. For $q > 1$, the planets will be finally locked into MMRs. As aforementioned, \citet{kley15} further showed that two circumbinary planets can be captured into MMR as a result of inward convergent migration. In the subsequent process, the planets usually undergo dynamical scattering \citep{kley15}.

In this work, we model the late evolution stage of circumbinary systems by ignoring the effect of the residual gas \citep{chatterjee08, juric08, raymond08, beauge12, moeckel12}. \citet{moeckel12} testified it is a reasonable approximation. The hydrodynamic outcomes of planet scattering in transitional discs are discussed in their work. They showed that N-body dynamics and hydrodynamics of scattering into one-planet final states are nearly identical. The eccentricity distributions of the surviving planets are almost unaltered by the existence of the residual gas.

The article is organized as follows. Section 2 presents our numerical model and initial conditions. In Section 3 we give the numerical results and the analyses. Section 4 discusses the probability of reproducing the orbital configurations of Kepler-34b and Kepler-314b. Finally, Section 5 summarizes the major results and discuss the orbital evolution theory of circumbinary planets.

\section{Model and initial conditions}
We start our scattering simulations with two planets which have been extensively studied in single star systems (see \citet{ford08} and the references therein). Another advantage of the two-planet model is that understanding this simple case facilitates the analysis of simulations with more planets. \citet{gong17} showed that binary configurations have no substantial effect on the scattering results. Therefore, we take a Kepler-16(AB)-like close binary configuration \citep{doyle11} as a baseline. That is, the SMA, eccentricity and the mass ratio of the binary are $a_{B}=0.22$ au, $e_{B}=0.16$, $q_{B}=M_{b}/M_{a}=0.29$, respectively. The total mass of the binary is 1 $M_{\odot}$.\footnote{The $a_{c}$ of the reference system is 0.634 au. The Kepler-16(AB) binary has a total mass of 0.87 $M_{\odot}$. Its $a_{c}$ is 0.635 au.} According to the mass distribution of the circumbinary planets (see Table 1), we consider five sets of mass combinations of the planets. We found that, in addition to the mass ratio, the initial relative position of the two planets also affects the simulation results. A bracket is used to indicate the initial relative position of the two planets. For example, [$M_{S}$, $M_{J}$] refers to the initial inner/outer planet is a Saturn-like/Jupiter-like planet. All the mass combinations are given in Table 2.

Among these \emph{Kepler} circumbinary planets, Kepler-1647b has the longest orbital period and is located at $7.4~a_{c}$ (2.7 au) \citep{kostov16}. The planet may not undergo significant disk-driven migration. Kepler-47c is the outer planet of a multiple-planet system. Except them, the majority of \emph{Kepler} circumbinary planets lie between $1.16~a_{c}$ and $1.85~a_{c}$. Hence, we use this interval to set the initial position of the inner planet. For the initial SMA of the inner planet, we consider two cases $a_{1,0}=1.2~a_{c}$ and $2.0~a_{c}$, respectively.  They represent the scattering occurring near the stable boundary of the binary, or away from it. The initial SMA of the outer planet is
\begin{equation}
a_{2,0}=a_{1,0}+KR_{Hill,m},
\end{equation} \label{a20}
where $R_{Hill,m}$ is the \emph{mutual} Hill radius defined as
\begin{equation}
{R_{Hill,\,m}} = {\left( {\frac{{{m_1} + {m_2}}}{{3{M_ * }}}} \right)^{{1 \mathord{\left/
 {\vphantom {1 3}} \right.
 \kern-\nulldelimiterspace} 3}}}\left( {\frac{{{a_1} + {a_2}}}{2}} \right).
\end{equation} \label{Rh}
$K$ is an important parameter which may affect the unstable timescale of the system \citep{kratter14, chatterjee08}.
A compromise strategy is taken in choosing the $K$ value in this work. We avoid unphysical very closely-packed systems (small $K$).  On the other hand, we do not take a large $K$ because the required integration time is too long to perform a large-sample statistical study.

For an isothermal and radiative disk, \citet{kley15} revealed that planets typically result in a capture of low-order MMR with period ratios of 3:2, 5:3, 2:1, etc. At $a_{1,0}=1.2~a_{c}$, we take $K=4$. The resultant initial distance of planet is larger than 3:2 and near 5:3 resonance ($K=3.4$ and 4.3, respectively) in our model. In a single star system, the unstable timescale of two-planet systems can be measured using Hill stability criteria \citep{gladman93}. For a [$M_{S}, M_{J}$] system with $a_{1,0}$=3 au in singe star system, the Hill stability criteria gives $K\sim3$ (the unit is the mutual Hill radius). Thus, in the binary system, we take $K=3$ for $a_{1,0}=2.0~a_{c}$ (scattering taking place away from the binary). Combined with our numerical tests, we integrate each system up to $10^6$ years for $a_{1,0}=1.2$, and $10^7$ years for $a_{1,0}=2.0~a_{c}$. Numerical test showed these integration times are long enough to reflect the scattering process. The initial eccentricities and inclinations of planets are $<10^{-3}$. All initial phase angles were assigned randomly from 0 to 2$\pi$. We fully integrated each system using the Bulirsch-Stoer integrator in our revised Mercury package \citep{chambers99}. For every mass combination and the different $a_{1,0}$ (see Table 2), we perform 1000 runs. The type is referred to as `ejections' meaning the distance between the planet and the barycenter of a binary is larger than 500 au.

\section{Numerical results}

In this work, we assume that the currently observed single-planet systems are the products of PPS of original multiple-planet systems. Thus, we focus on the resulting single-planet systems. Some of them are the merger of the two planets. Considering the mass and angular momentum conservation, the new planet generally has a larger mass and a low eccentricity, which is similar to the results of PPS in single star systems \citep{ford01, ford08}. The majority of single-planet systems come from a scenario in which one planet is ejected out of the system. In the following, we analyze these systems in detail.

\subsection{Mass ratio vs. ejection preference}

In single star systems, the ejections are of the less massive one of the two planets, regardless of whether it was initially the inner or the outer one \citep{ford08}. For the circumbinary planets, this conclusion is conditional. It depends on the planet mass ratio $q$ ($<1$) and the initial relative location of the two planets (see Table 2). \citet{gong17} found that for equal-mass planets, the initial inner planets are peculiarly prone to be ejected if PPS takes place near the unstable boundary of the binary. This trend is maintained as long as the mass ratio of the planets is greater than 0.3 as we can see in Table 2 where $a_{1,0} = 1.2~a_{c}$.

However, as the mass ratio becomes smaller, this tendency disappears. For $q$ = 0.03, the less massive planets are always scattered out of the system, regardless of its initial relative position. Our numerical simulation suggested if the planetary mass ratio is greater than $\sim 0.3$, the initial inner planets are more likely to be ejected. However, if the mass ratio is less than 0.3, the less massive planets are easily to be scattered out of the system. This position dependency does not exist if the initial locations of the two planets are moved away from the binary ($a_{1,0} = 2.0~a_{c}$). Regardless of its initial position, the ejections are of the less massive planets.

Kepler-34b and Kepler-413b have a mass of $\sim 0.2M_{J}$. We carried out additional simulations to explore how the total mass of two planets affects above results. We set the mass of the more massive planet to be $0.2M_{J}$, and four mass ratios $q=1, 0.5, 0.3, 0.15, 0.03$ were considered. In addition to $K$, the initial distance between two planets is also relevant to their total mass (see Equation 2). Through numerical examination, we adopted $K=5$ to avoid very closely-packed system. The other parameters remain unchanged. The results for $a_{1,0}=1.2a_{c}$ are shown in Table 3. As can be seen from Table 3, although the fraction is slightly different, the general trend of the ejection preferences is similar to each other. It indicates that the total mass has little effect on the outcomes, at least for giant planets with a mass $m_{p} \approx 0.2M_{J}$.

As aforementioned, several studies have shown that planets born in a circumbinary disk will migrate inward and eventually be stalled near the inner hole of the disk \citep{pierens08,kley15}. If an outer planet migrate toward it, PPS will occur. If we assume that the mass-power-law of the circumbinary planets formed in a system is the same as that of  the Solar System, the inner planet (Jupiter) is more massive than the outer planet (Saturn). Thus, our results imply there is an equivalent or even a larger probability for the inner more massive planets to be ejected out of the system, if their mass ratio is larger than a critical value. The currently observed \emph{Kepler} circumbinary planets are generally less massive. Whether PPS occurring in the late stage is a possible mechanism accounting for this phenomenon, which should be examined by future observations and investigations.

\subsection{Mass ratio vs. orbital element distribution}
We discuss the final orbital distribution of the surviving planets in two subsets ($a_{1,0} = 1.2~a_{c}$ and $a_{1,0} = 2.0~a_{c}$). Figure 1 shows the eccentricity distributions of the surviving planets for $a_{1,0} = 1.2~a_{c}$. To show the details, we present the results according to different mass ratios and initial relative positions. On the top panel of Figure 1, the final eccentricity distributions are of the initial inner planets which survive the PPS. The bottom panel shows the eccentricity distribution of the initial outer planets which survive PPS. For clarity, we only draw the cases of $q$ = 0.3 and 0.03. From Figure 1, we find that if the mass ratio of the planets is small $q$ = 0.03, the remaining planets maintain a small eccentricity. The values are roughly equivalent to their forced eccentricities. This is because in the scattering process, the less massive planet was scattered out of the system quickly under the combined actions of the massive one and the binary. As a result, the more massive planet gets little angular momentum, so it maintains a small eccentricity. An example is shown in Figure 2.

For $q$ = 0.3, to gain a lower eccentricity, the surviving planets must be the massive one of two initial planets, regardless of which planet was initially closer (see the red and black line on the top and bottom panel, respectively). Conversely, when the surviving planets are the less massive one, their eccentricities are generally large with a median value 0.6. A case is given in Figure 3. Besides, the range of eccentricity is related to the initial relative position of the less massive planet. If it is the initial inner one, their final eccentricities are larger than the initial outer one, indicating that it gets more angular momentum in a more violent dynamical process. It seems that the eccentricity distribution of the outer surviving less massive planet is more diffuse than the inner one. If we assume that the currently observed circumbinary planet is the product of PPS, the value of its eccentricity can be used to estimate the mass ratio of the original two planets.

\citet{gong17} found that, after PPS, the SMA of the surviving planets increase contrary to the scattering phenomena in single star systems. In order to study the SMA variation of the remaining planets, we plot the distribution of $a_{p}/a_{p,0}$ (the ratio of the final SMA of the surviving planet to its initial SMA) in Figure 4. If the ratio is greater than 1, the planet is shifted outward after PPS. We find that, nearly in all cases, the SMAs of the surviving planets are $>1$ for $a_{1,0} = 1.2~a_{c}$ cases. The results are not related to the planet mass ratio and the initial relative position of the two planets. For $q$ = 0.3, the maximum $a_{p}/a_{p,0}$ can reach $\sim15$.

For $a_{1,0} = 2.0~a_{c}$, the eccentricity distribution does not have significant change (see Figure 5). But in the SMA distribution a double peak structure emerges as we can see in Figure 6. It means that some of the surviving planets migrated inward during PPS, which is a typical feature of the scattering in single star systems. It indicates that at $2.0~a_{c}$, scattering begins to appear the characteristics of PPS in single star system. The influence of the inner binary is weakened at this location. In Figure 7, we show that the SMA of a surviving planet shrinks after PPS. Our additional numerical simulations showed that if the scattering occurs in the region of [1.2---1.8]$a_{c}$, the scope of most currently discovered \emph{Kepler} circumbinary planets, almost all the SMAs of the surviving planets are incremented after PPS. It suggests that if these planets were the survivors of PPS, their initial location would have been closer than their currently observed value.

Moreover, we found that the surviving planet generally maintains a nearly coplanar configuration with the binary. Besides, the distribution of the inclination is also related to the mass ratio and the initial relative position of the two planets. The inclination distribution of the surviving planets is related to the initial inclination of the two planets, which we will discuss in the future.

\section{Application to Kepler-34b and Kepler-413b}

\subsection{Kepler-34b}
The orbital evolution of Kepler-34b in a protoplanetary disc has been studied in several works. \citet{pierens13} considered the migration and gas accretion scenarios of Kepler-34b. For the fully-formed planet, its the final location is determined by the adopted physical parameters of the disc (aspect ratios $h$, viscous stress parameters $\alpha$, etc.). In addition, they took into account low-mass cores that migrate and accrete gas while the disc is being dispersed. In most cases, the planets halt migration beyond its currently observed orbit (1.09 au). However, in the case of $h=0.05$ and $\alpha=10^{-4}$, the planets can migrate across its currently observed orbit and reach $\sim$1 au with a nearly zero eccentricity. Recently, \citet{mutter17} modeled the disc self-gravity in sculpturing the structure of circumbinary discs. They showed the scale of the inner cavity depends on the disc mass. An enhanced disk mass will cause the outer edge of the cavity closer to the binary. It may imply that the circumbinary planet formed in the disks can migrate into much closer region.

Using a more sophisticated disc model, \citet{kley15} revisited the evolution of Kepler-34b. In their work, they further indicated that the planet stalls beyond the observed regime of Kepler-34b. To account for the closer orbit of Kepler-34b, they modeled a two-planet scenario in the simulations. They showed that the two planets can enter a 3:2 MMR and then undergo a sequential weak scattering events. As a result, the inner planet can move toward the present orbit of Kepler-34b. The model may imply that there is an additional planet in the system ($\sim 1.5$ au or 1.8 $a_{c}$). However, in our model, we assume that the currently observed single-planet system could be simply the survival of PPS of an original two-planet system. After PPS, one planet is completely ejected out of the system. One of our key goals herein is to observe whether this model can reproduce the current orbital configuration of Kepler-34b. In the following, we will discuss this scenario.

As we mentioned, if the planets migrate inward and stall beyond the presently observed orbit of Kepler-34b, the ejection model cannot reproduce its current orbital configuration. After PPS, the surviving planet will migrate \emph{outwards}, leading to a larger SMA. However, if Kepler-34b ever migrated to a closer location as discussed in \citet{pierens13}, PPS can reproduce the observed orbital configuration of Kepler-34b.

We set the initial orbital elements of the inner planets according to \citet{pierens13} (for $h=0.05$ and $\alpha=10^{-4}$), where $a_{1,0}\approx$1 au, $e_{1,0}\approx$0.01. The SMA of the outer planet is given according to the 3:2 MMR region with respect to the inner planet. We give the eccentricity of the outer planets an estimated value referring to the other case in \citet{pierens13}, but our result does not depend on this value. Then, we performed a number of simulations, where one of the runs is shown in Figure 8. From Figure 8, we show that two planets \emph{ejection} model can reproduce the observed orbital configuration of Kepler-34b through the disk-driven migration plus PPS model. Using the given parameters and different phase angles, we performed 1000 runs. Among 399 systems, one planet was ejected out of the system and the other finally survived, whereas in 126 systems two planets merged to form one large planet. In 18 systems, no planet remained. The remaining systems had two planets at the end of the integration. We found that in 20 systems the surviving planet had a similar eccentricity and SMA (with an error bar) of Kepler-34b. Therefore the percentage of producing a Kepler-34b-like planet in our model is $20/399 \approx 5.01\%$.

\subsection{Kepler-413b}
Keple-413(AB) is a \emph{K}+\emph{M} eclipsing binary with $a_{B}=0.101$ au \citep{kostov14}. The mass of the two dwarfs are $m_{A}=0.82 M_{\odot}$ and $m_{B}=0.54 M_{\odot}$, respectively. A Neptune-size circumbinary planet, Kepler-413b, was discovered in this system on an eccentric orbit with $a_{p}=0.355$ au, $e_{p}=0.12$.  At present, there is no disc migration model of Kepler-413b. Hydrodynamical simulations showed that the eccentricity of the binary has a decisive influence on the size and structure of the disk's inner cavity, and the final position of a planet depends on this size. For
binary systems with nearly circular orbits, planets forming farther out in the calmer environment of the disk can migrate toward the unstable boundary of the binary \citep{kley14}. The Kepler-413(AB) binary has a small eccentricity ($e_{B}=0.037$). It seems unlikely that Kepler-413(AB) can open a wide inner hole in the disk as the highly eccentric binary Keple-34(AB) with $e_{B}=0.52$ \citep{kley15}. Thus, we assume the planet born in Kepler-413(AB) system can migrate toward the innermost region of the disk, like Kepler-16b ($\sim 1.2~a_{c}$).

Herein we take $a_{1,0}=1.2~a_{c}$. Similarly, the initial orbit of the outer planet is set according to the 3:2 MMR location. Then, we carried out a set of simulations to investigate the orbital configuration of the Kepler-413b. A typical case is illustrated in Figure 9. The initial orbital parameters of two planets are $a_{1,0}=0.31$ au, $e_{1,0}=0.07$, $a_{2,0}=0.406$ au and $e_{2,0}=0.02$, respectively. The masses of two planets are $m_{inner}=0.211~M_{J}$ (Kepler-413b) and $m_{outer}=0.09~M_{J}$, respectively. At $\sim 2\times10^{6}$ yr, the outer less massive planet was ejected out of the system. From the simulations, we found that two planets \emph{ejection} model can be at work to generate the final orbital configuration of Kepler-413b as currently observed.  This suggests that both the eccentricity and SMA of Kepler-413b can be well reproduced in two-planet ejection model. Using the given parameters, we carried out 1000 runs for this system by varying the phase angles. Among 422 runs, one planet was ejected and the other planet survived in the system, while for 211 systems two planets merged into one planet. In 16 systems there was no planet left. The remaining systems were observed to occupy two planets when the simulations were done. We found that in 41 systems the surviving planet bears a similar eccentricity and SMA of Kepler-413b. Hence, we have come a conclusion that the percentage of yielding a Kepler-34b-like plane is $41/422 \approx 9.72\%$ in our ejection model.

\section{Summary and discussion}

As known, PPS scenarios can shed light on several observational features of exoplanets, such as the formation of hot-Jupiters \citep{rasio96, nagasawa11, beauge12}, the stellar obliquity distribution of stars with hot-Jupiters \citep{winn15}, the eccentricity distribution of giant planets \citep{chatterjee08, juric08}. Furthermore, recent studies have shown that some of the circumbinary planets are likely to have undergone scattering process. In a single-stellar system, the scattering model can be at work to reproduce the eccentricity distribution of extrasolar giant planets, and the mass ratio of the planets is a vital parameter to understand their dynamical evolution.

In the present work, we concentrate on investigating the role of the mass ratio of the circumbinary planets over the scattering results. We first assume that the currently observed single-planet system is the survivor of PPS of an original multi-planetary system. Next, we extensively explore the effect of the mass ratio on the ejection preference and the orbital distribution of the surviving planets. Our simulations showed that the binary is involved in the scattering scenario, which makes the scattering results greatly different from those of PPS in single star systems. In addition, combined with the disk-driven models of circumbinary planet, we have studied the orbital configuration formation of Kepler-34b and Kepler-413b based on a planet-planet ejection scenario. Thus, we summarize the major conclusions of this work as follows.

1. Ejection preference is related to the planetary mass ratio and the scattering location. If the mass ratio of the two planets is greater than a critical value ($\sim 0.3$ in our model), the inner planet has an equivalent or even larger probability to be ejected out of the system as the PPS takes place nearby the unstable boundary of the binary. If the mass ratio is less than the critical value or the scattering position is moved away from the binary, the ejections are always the less massive planets.

2. The eccentricity distribution of the surviving planets varies with the mass ratio and the initial relative position of the two planets. To obtain a low eccentricity, the surviving planet has to be the more massive one, regardless of its initial location (inner or outer). If the mass ratio of the planets is less than $\sim 0.3$, the remaining planets can maintain a small eccentricity, which is nearly equal to its forced eccentricity.

3. Within the range of [1.2---1.8]
$a_{c}$, the SMA of the surviving planets always increase after PPS. If the innermost region that a planet can reach (driven by the disk) is beyond its currently observed location, the two-planet ejection model cannot reproduce their current orbital configuration.

4. In the migration and ejection scenario, the formation of the configuration of Kepler-34b or Kepler-413b seems to be likely from our simulations. It requires  that the planets beforehand migrated closer to the binary as indicated in \citet{pierens13}. Its universality needs a more mature disk model to elucidate  in the future.

Compared to the protoplanetary disk in a single star system, the architecture of the circumbinary disk seems to be more complicated \citep{fleming17}. The evolution of the disk is relevant to the mass ratio of the binary, its eccentricity and the physical parameters of the circumbinary disk. \citet{thun17} studied how above factors affect the gap size of the disk. Interestingly, they found that there is a bifurcation occurring at around $e_{B} \approx 0.18$ where the gap is smallest. For $e_{B}$ smaller and larger than this value, the gap size can increase. It is worth further investigating how this feature of circumbinary disk plays a role in the orbital evolution of a formed planet. In the meantime, planetary accretion, growth and migration should be considered in the context of the physical evolution of the disk. In particular, the final location of a circumbinary planet is determined by a delicate interplay between the detailed structure of the tidal-formed cavity and the orbital parameters of the planet, the dissipation of the disk, the disc self-gravity, etc. Actually, these are open questions for the planetary community. However, the disk-driven migration of circumbinary planets and the subsequent PPS make it possible to shape the final orbital configuration of \emph{Kepler} circumbinary planets, and the detailed scenarios should be investigated in the forthcoming study.

\acknowledgments
We thank the anonymous referee for constructive comments and suggestions to improve the manuscript. This work is financially supported by National Natural Science Foundation of China (Grants No. 11773081, 11573018, 11473073, 11661161013), the Strategic Priority Research Program-The Emergence of Cosmological Structures of the Chinese Academy of Sciences (Grant No. XDB09000000), the innovative and interdisciplinary program by CAS (Grant No. KJZD-EW-Z001), the Natural Science Foundation of Jiangsu Province (Grant No. BK20141509) and the Foundation of Minor Planets of Purple Mountain Observatory. Gong Y.-X. also acknowledges the support from Shandong Provincial Natural Science Foundation, China (ZR2014JL004).

\clearpage
\clearpage

\begin{table}
\begin{center}
\caption{Mass and Orbital Configuration of \emph{Kepler} Circumbinary Planets. \label{tbl-1}}
\begin{tabular}{lccrcc}
\tableline
\tableline
Planet Name  &    Mass ($M_{J}$)  & Semimajor Axis (au) & Eccentricity  & Forced Eeecentricity$^{(a)}$ & $a_{p}/a_{c}$  \\
\tableline
Kepler-16b        &0.33   & 0.705    & 0.007  &  0.034   & 1.16   \\
Kepler-34b        &0.22   & 1.090    & 0.182  &   0.002       & 1.31    \\
Kepler-35b        &0.13   & 0.603    & 0.042  &  0.002        & 1.21   \\
Kepler-38b        &0.38   & 0.464     & $<$0.032    &  0.024        &  1.25  \\
Kepler-47b        &---   & 0.296    & $<$0.035     & 0.004   &  1.48  \\
Kepler-47c        &---   & 0.989     & $<$0.411    & ---     &  4.95\\
Kepler-64b        &0.53  & 0.643     & 0.054    & 0.044    & 1.26  \\
Kepler-413b       &0.21   & 0.355    & 0.118   & 0.003 & 1.40  \\
Kepler-453b       &0.03   & 0.788    & 0.038  & --- & 1.85  \\
Kepler-1647b      &1.52   & 2.721    & 0.058  & ---& 7.41  \\
\tableline
\end{tabular}
\tablenotetext{}{$^{a}$ \citet{bromley15}}
\tablecomments{Data in columns 2---4 are taken from \it{http://exoplanet.eu/}.}
\end{center}
\end{table}

\clearpage

\begin{table}
\begin{center}
\caption{Ejection Preference of the Two Circumbinary Planets.\label{tbl-2}}
\begin{tabular}{l|l|cc|cc|cc}
\tableline
\tableline
Mass ratio & Planets  &\multicolumn{2}{|c|}{1.2$a_{c}$}& \multicolumn{2}{|c|}{2.0$a_{c}$} & \multicolumn{2}{|c}{Single-star System$^{(a)}$}   \\
\tableline
\multicolumn{2}{l|}{Ejection Preference}  & Eje$_{in}$ & Eje$_{out}$ & Eje$_{in}$  & Eje$_{out}$ & Eje$_{in}$ &  Eje$_{out}$ \\
\tableline
 $q=1$ & $[M_{J}, M_{J}]$        & 0.39  & 0.15    & 0.25  &  0.19        & ---   & ---\\
 \tableline
 $q=0.5$ & $[0.5M_{J}, M_{J}]$     & 0.46  & 0.08    & 0.42  &  0.13        & ---   & ---\\
       &  $[M_{J}, 0.5M_{J}]$     &0.36   & 0.21     & 0.19    &  0.33        &  ---  & ---\\
  \tableline
 $q=0.3$ & $[M_{S}, M_{J}]$        &0.44   & 0.03    & 0.62     & 0.01   &  0.13  & 0.00\\
         &  $[M_{J}, M_{S}]$        &0.24   & 0.22     &0.03   & 0.54     &  0.00& 0.17\\
  \tableline
 $q=0.15$ & $[0.5M_{S}, M_{J}]$        &0.45  & 0.01     & 0.54  & 0.00    & ---  & ---\\
          &  $[M_{J}, 0.5M_{S}]$       &0.12   & 0.35    & 0.00 & 0.53 & ---  & ---\\
  \tableline
 $q=0.03$ & $[0.1M_{S}, M_{J}]$        &0.65  & 0.00     & 0.53  & 0.00    & ---  & ---\\
          &  $[M_{J}, 0.1M_{S}]$       &0.00   & 0.69    & 0.00 & 0.52 & ---  & ---\\
\tableline
\end{tabular}
\tablenotetext{}{$^{a}$ A set of PPS simulation (1000 runs) in single star systems is performed for comparison. The initial SMAs of the two planets are $a_{1, 0}=3$ au, $a_{2, 0}= a_{1, 0} + KR_{Hill, m}$, $K=3$. The initial distance between the two planets is close to their Hill stability boundary.}
\tablecomments{Eje$_{in}$ (Eje$_{out}$) is the fraction that the initial inner (outer) planets were ejected out of the systems in total 1000 runs.}
\end{center}
\end{table}

\clearpage

\begin{table}
\begin{center}
\caption{Ejection Preference of Two Less Massive Planets.\label{tbl-3}}
\begin{tabular}{l|l|cc}
\tableline
\tableline
Mass ratio & Planets  &\multicolumn{2}{|c}{1.2$a_{c}$}\\
\tableline
\multicolumn{2}{l|}{Ejection Preference}  & Eje$_{in}$ & Eje$_{out}$  \\
\tableline
 $q=1$ & $[0.2M_{J}, 0.2M_{J}]$        & 0.52  & 0.19    \\
 \tableline
 $q=0.5$ & $[0.1M_{J}, 0.2M_{J}]$     & 0.55  & 0.06    \\
       &  $[0.2M_{J}, 0.1M_{J}]$     &0.37   & 0.26     \\
  \tableline
 $q=0.3$ & $[0.06M_{J}, 0.2M_{J}]$     & 0.69  & 0.02    \\
       &  $[0.2M_{J}, 0.06M_{J}]$     &0.35   & 0.32     \\
  \tableline
 $q=0.15$ & $[0.03M_{J}, 0.2M_{J}]$        &0.72   & 0.01    \\
         &  $[0.2M_{J}, 0.03M_{J}]$        &0.29   & 0.41     \\
  \tableline
 $q=0.03$ & $[0.006M_{J}, 0.2M_{J}]$        &0.70   & 0.00    \\
         &  $[0.2M_{J}, 0.006M_{J}]$        &0.00   & 0.69     \\
\tableline
\end{tabular}
\end{center}
\end{table}

\clearpage

\begin{figure}
\epsscale{0.8} \plotone{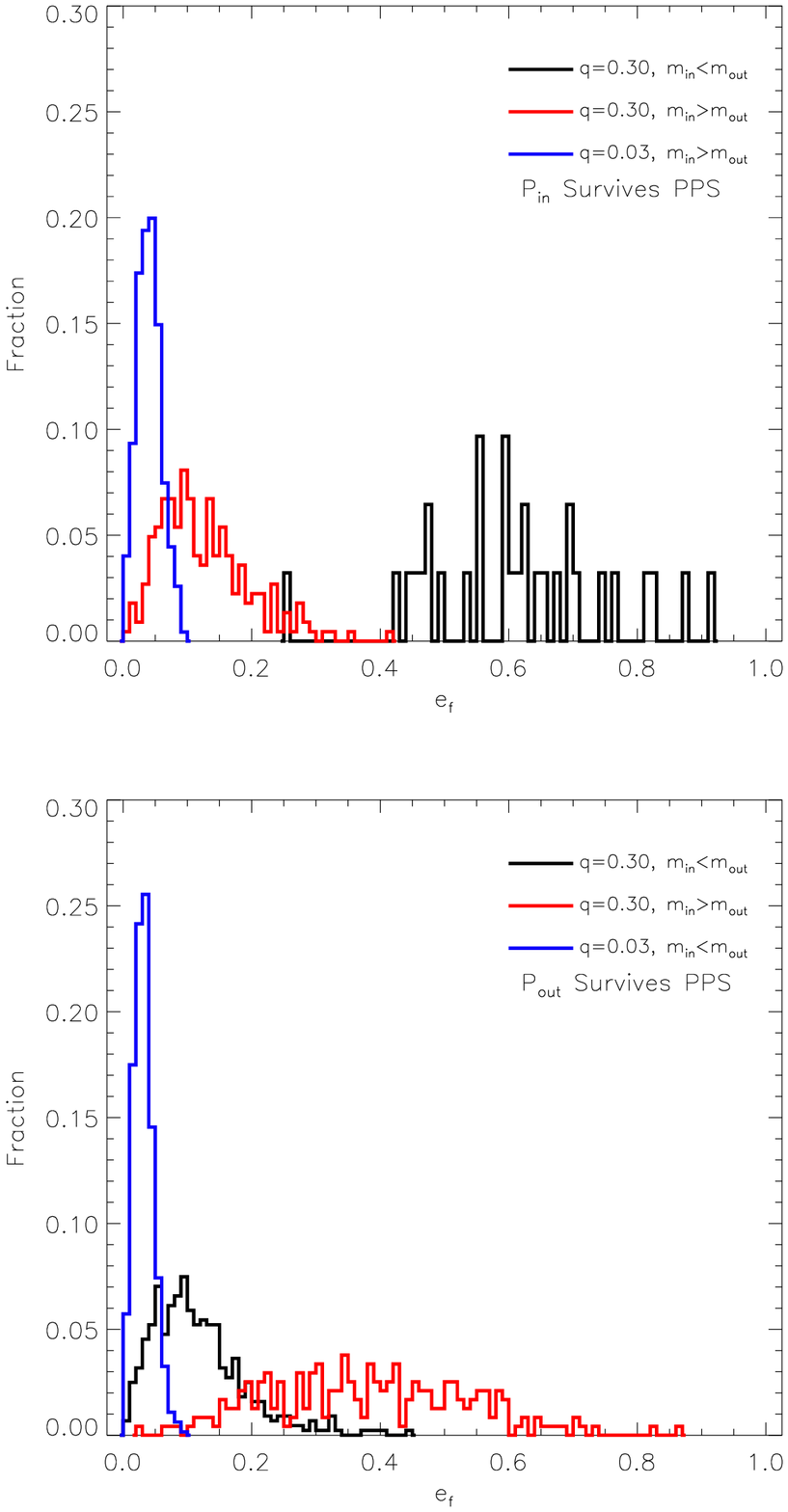} \caption{The eccentricity distribution of the surviving planets after one planet was ejected out of the system. On the top panel, the surviving planets were the initial inner planets. The bottom panel shows the eccentricity distribution of the initial outer planets which survived PPS. Different line represents different mass ratio and initial relative position of the two planets. The initial semi-major axis of the inner planets are 1.2 $a_{c}$.
\label{fig1}}\end{figure} \clearpage

\begin{figure}
\epsscale{1.0} \plotone{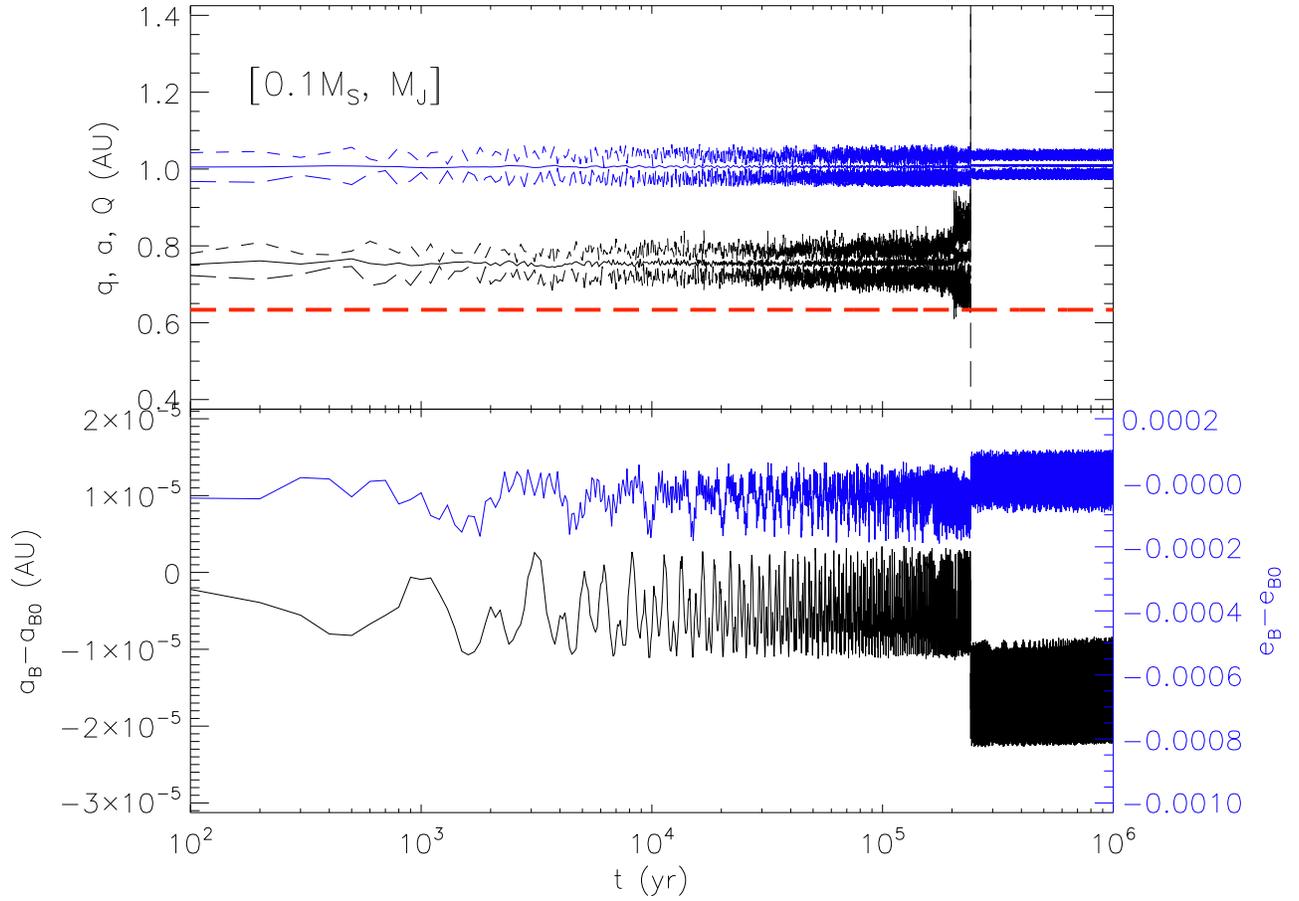} \caption{Upper panel: time evolution of SMA (a), pericenter (q) and apocenter (Q) distances of two planets.  The initial inner planet was ejected out of the system at $\sim 2\times10^{5}$ yr. The final eccentricity of the surviving planet was $\sim 0.03$. The dashed red line denotes the corresponding $a_{c}$ derived by \citet{holman99}. Lower panel: time evolution of the semi-major axis ($a_{B}-a_{B,0}$) and eccentricity ($e_{B}-e_{B,0}$) of the inner binary.
\label{fig2}}\end{figure} \clearpage

\begin{figure}
\epsscale{1.0} \plotone{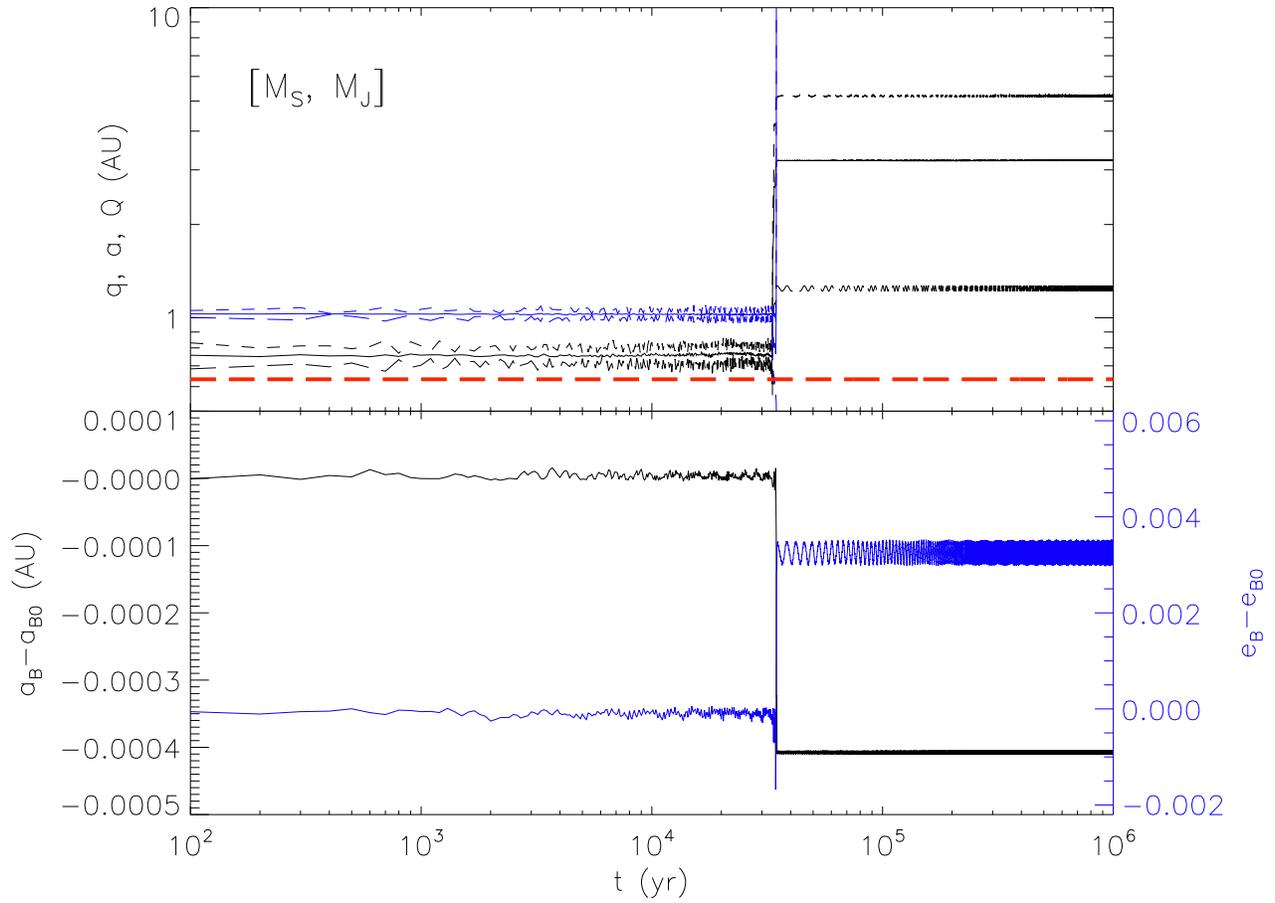} \caption{Conventions are as in Figure 2. The initial more massive planet was ejected out of the system at $\sim 3.5\times10^{4}$ yr. As a result, the surviving planet got a high eccentricity of 0.6.
\label{fig3}}\end{figure} \clearpage

\begin{figure}
\epsscale{0.8} \plotone{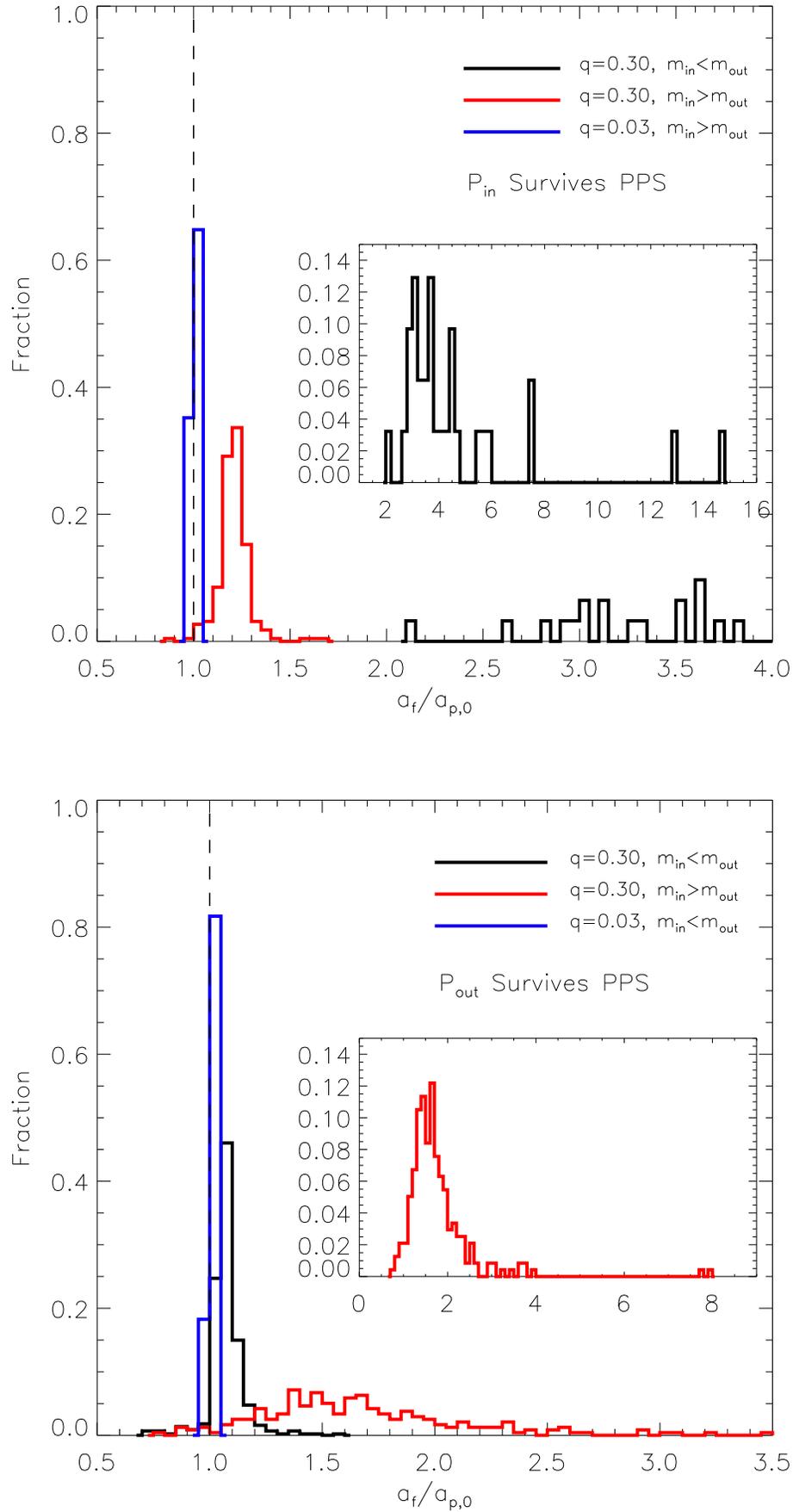} \caption{Final SMA distributions of the surviving planets after PPS. The dashed lines are for $a_{f}/a_{p,0}$=1. The subgraph in the top and bottom panel shows the full range distribution of the same color line but with different bin size. Other conventions are as in Figure 1.
\label{fig4}}\end{figure} \clearpage

\begin{figure}
\epsscale{0.8} \plotone{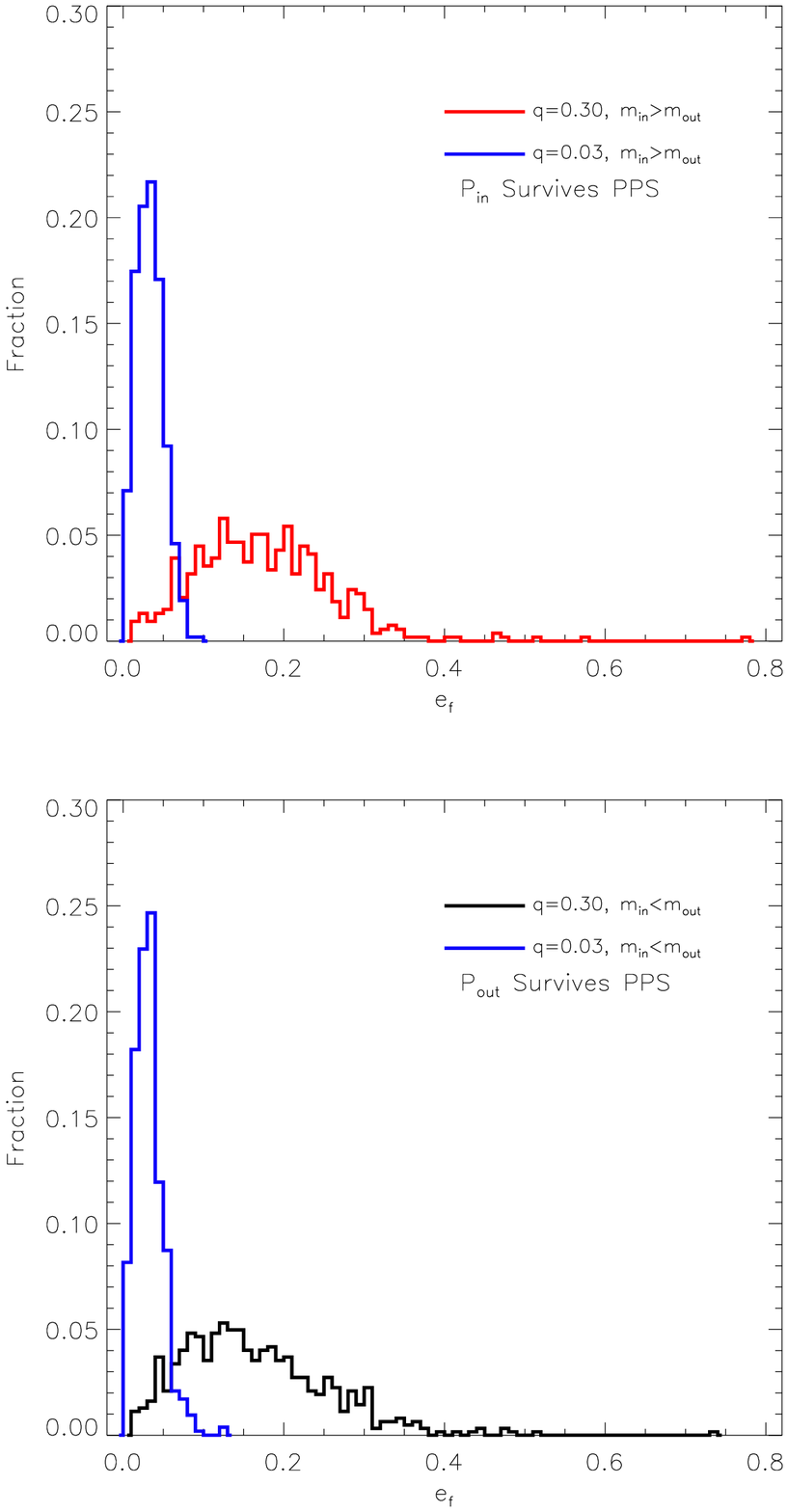} \caption{Final eccentricity
distribution of the surviving planets for $a_{1,0}=2.0a_{c}$. Conventions are as in Figure 1.
\label{fig5}}\end{figure} \clearpage

\begin{figure}
\epsscale{0.8} \plotone{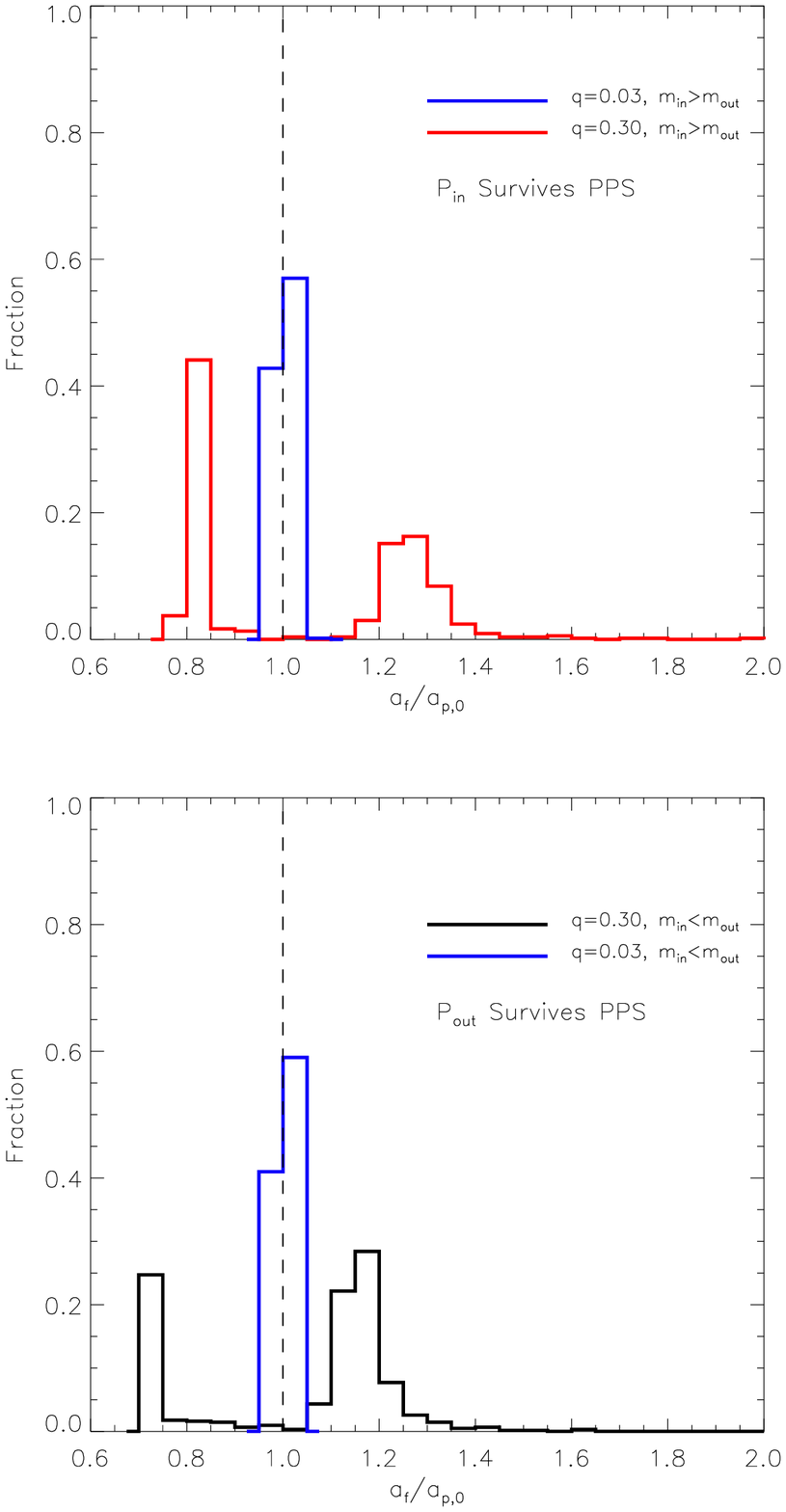} \caption{Final SMA
distribution of the surviving planets for $a_{1,0}=2.0a_{c}$. Conventions are as in Figure 4.
\label{fig6}}\end{figure} \clearpage

\begin{figure}
\epsscale{1.0} \plotone{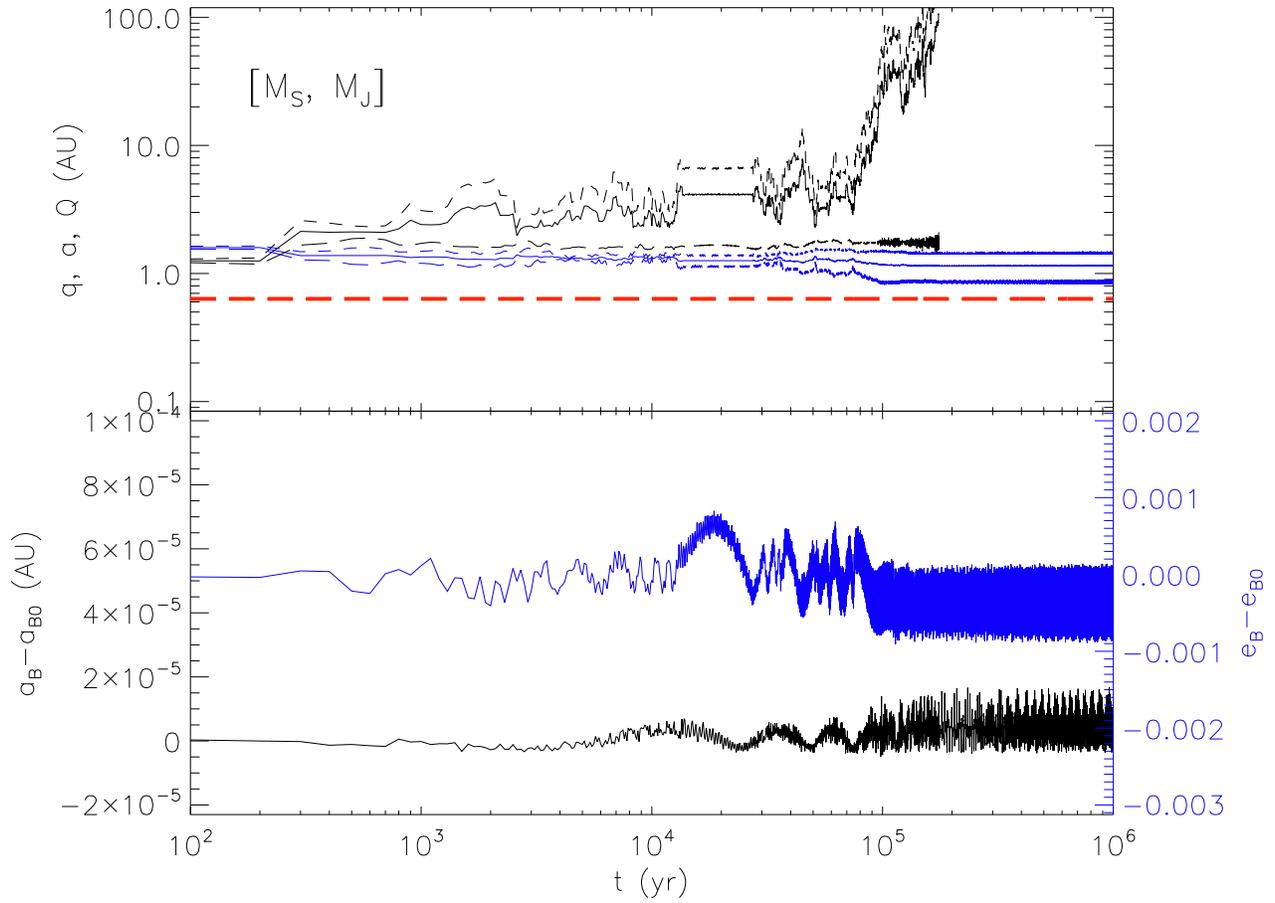} \caption{Conventions are as in Figure 2. The initial inner planet was ejected out of the system at $\sim 2\times10^{5}$ yr. The SMA of the initial outer planet shrank after PPS.
\label{fig7}}\end{figure} \clearpage

\begin{figure}
\epsscale{1.0} \plotone{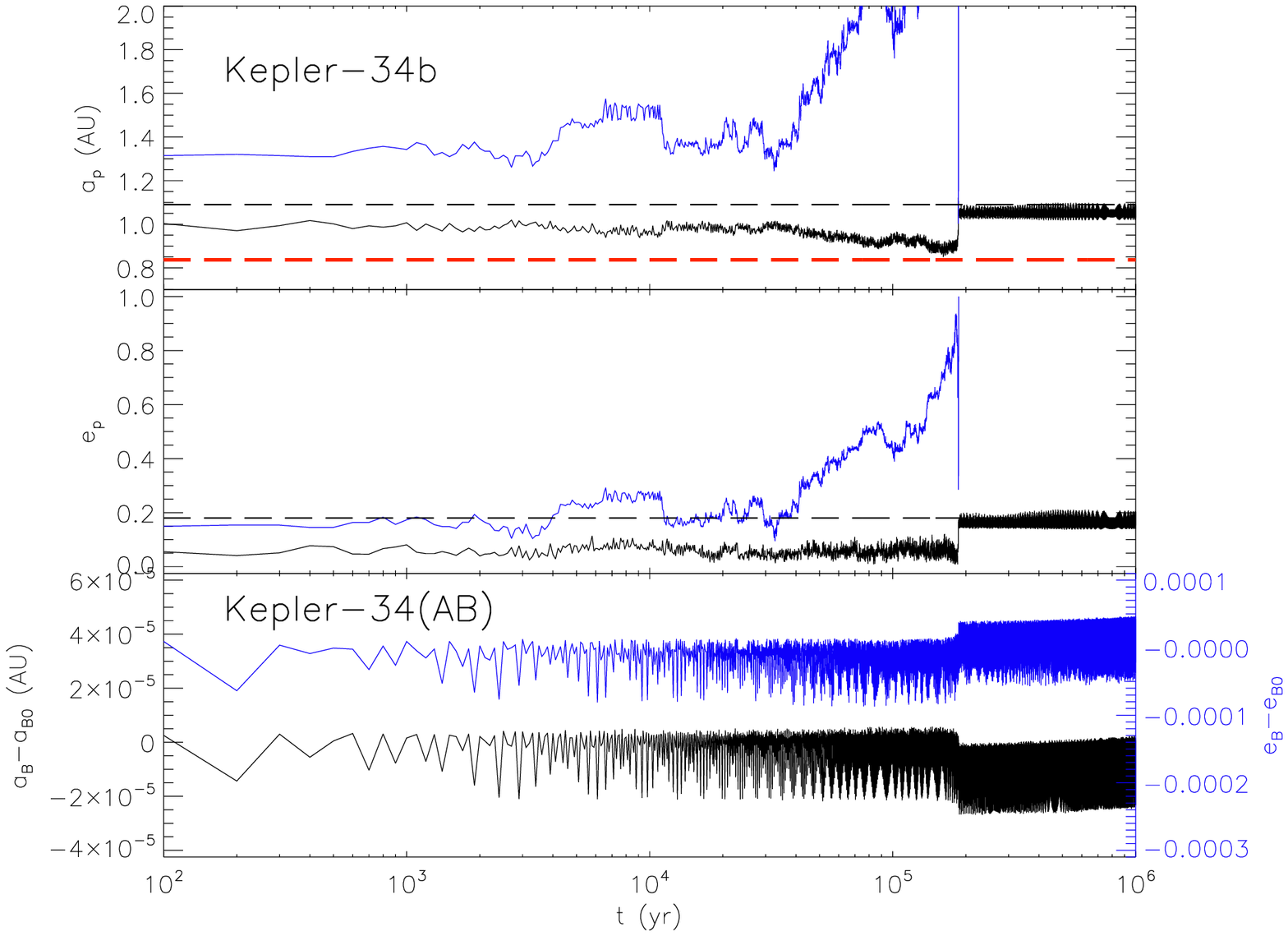} \caption{Two planets \emph{ejection} model reproduces the observed orbital configuration of Kepler-34b. The binary is Kepler-34(AB) \citep{welsh12}. Its orbital parameters and the masses are $a_{B}=0.229$ au, $e_{B}=0.521$, $m_{A}=1.048 M_{\odot}$, $m_{B}=1.021 M_{\odot}$. The initial orbital parameters of the two planets are $a_{1,0}=1$ au, $e_{1,0}=0.02$, $a_{2,0}=1.31$ au, $e_{2,0}=0.15$. The initial SMA of the outer planet is given by the location of 3:2 MMR with the inner one. The mass of the two planet are $m_{inner}=0.22M_{J}$ (Kepler-34b) and $m_{outer}=0.06M_{J}$, respectively. The evolution of the SMAs and eccentricities of the two planets are shown in top and middle panel. The evolution of SMA and eccentricity of the binary are plotted in the bottom panel. The dashed black lines in the top and middle panel represent the observed value. The dashed red line denotes the corresponding $a_{c}$ derived by \citet{holman99}.
\label{fig8}}\end{figure} \clearpage

\begin{figure}
\epsscale{1.0} \plotone{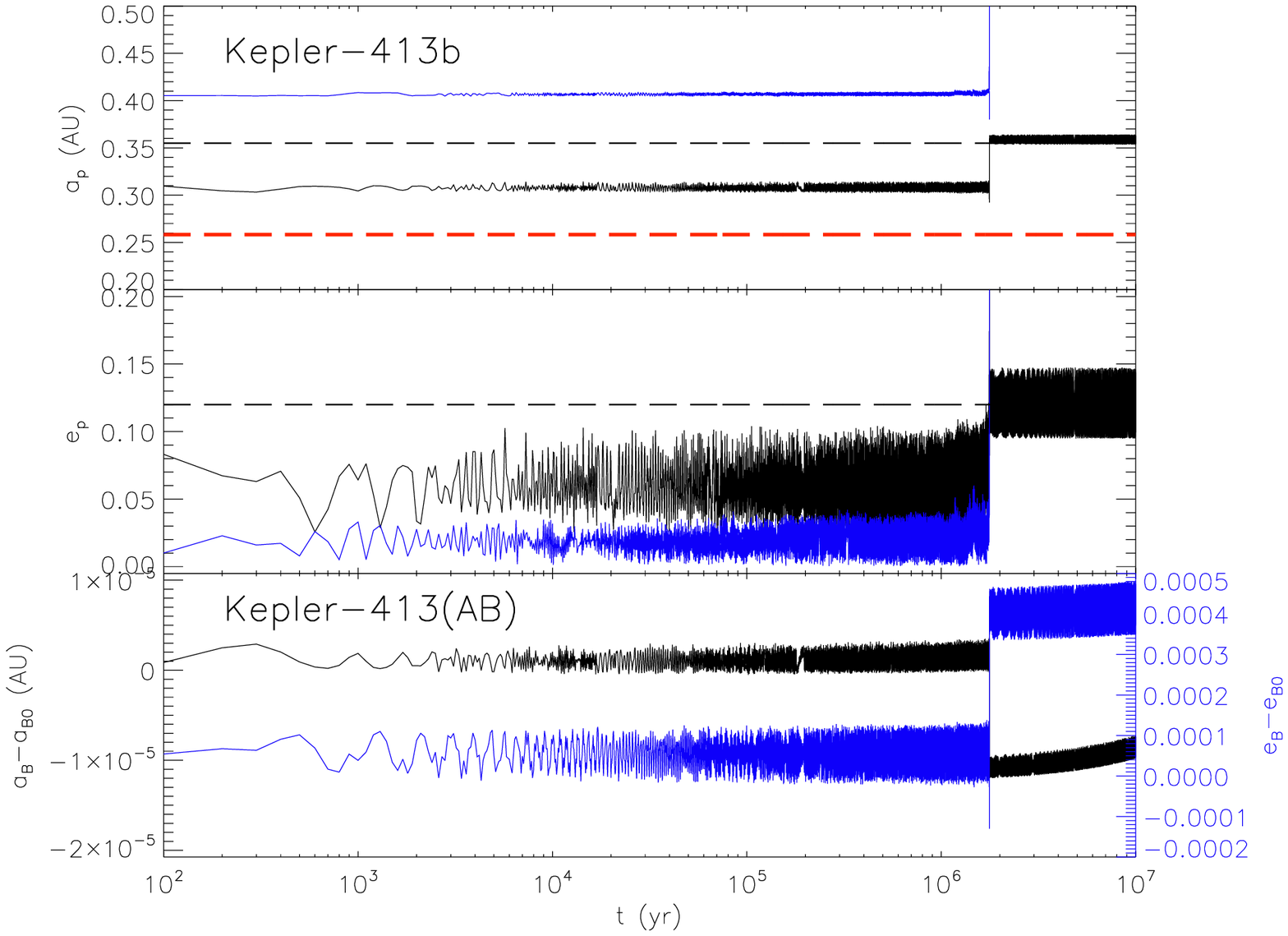} \caption{Two planets \emph{ejection} model reproduces the observed orbital configuration of Kepler-413b. The binary is Kepler-413(AB) \citep{kostov14}. Its orbital parameters and the masses are $a_{B}=0.101$ au, $e_{B}=0.037$, $m_{A}=0.82 M_{\odot}$, $m_{B}=0.54 M_{\odot}$. The initial orbital parameters of the two planets are $a_{1,0}=0.31$ au, $e_{1,0}=0.07$, $a_{2,0}=0.406$ au, $e_{2,0}=0.02$. The initial SMA of the outer planet is given by the location of 3:2 MMR with the inner one. The mass of the two planets are $m_{inner}=0.211M_{J}$ (Kepler-413b) and $m_{outer}=0.09M_{J}$, respectively. The evolutions of the SMAs and eccentricities of the two planets are shown in the top and middle panel. The evolution of the SMA and eccentricity of the binary are plotted in the bottom panel.
\label{fig9}}\end{figure} \clearpage

\end{document}